\documentclass[%
 preprint,
superscriptaddress,
amsmath,amssymb,
aip,
longbibliography,
]{revtex4-2}

\usepackage[utf8]{inputenc}
\usepackage[T1]{fontenc}
\usepackage{mathptmx}
\usepackage{etoolbox}
\usepackage{booktabs}

\usepackage{graphicx}
\usepackage{dcolumn}
\usepackage{bm}

\begin{document}


\title{DWS-based microrheology of triblock copolymers}

\author{Ren\'{e} Tammen}
 \affiliation{ 
PoreLab and Department of Physics, The Norwegian University of Science and Technology (NTNU), Trondheim NO-7491, Norway
}%

\author{Xiaoying Tang}%
\author{Ren Liu}%

\affiliation{ 
Cavendish Laboratory, University of Cambridge, Cambridge CB3 0HE, United Kingdom
}%

\author{Iliya D. Stoev}
\affiliation{%
Institute of Biological and Chemical Systems - Functional Molecular Systems, Karlsruhe Institute of Technology, Eggenstein-Leopoldshafen 76344, Germany
}%

\author{Erika Eiser}
\affiliation{
PoreLab and Department of Physics, The Norwegian University of Science and Technology (NTNU), Trondheim NO-7491, Norway
}
\affiliation{
Cavendish Laboratory, University of Cambridge, Cambridge CB3 0HE, United Kingdom
}

\email{erika.eiser@ntnu.no}

\date{\today}

\begin{abstract}
The thermally reversible phase transitions in aqueous solutions of the triblock copolymers known as Pluronic\textsuperscript{\tiny\textregistered} and their related textures are well-researched. However, their corresponding rheological properties are less studied. In particular, their high-temperature behavior is difficult to access with classical rheology. Here we demonstrated that Diffusing Wave Spectroscopy (DWS)-based microrheology allows us to study the phase transition and the associated viscoelastic properties of Pluronic\textsuperscript{\tiny\textregistered} F127 solutions for temperatures from 5$\,^{\circ}$C to 80$\,^{\circ}$C. From the measured intensity-autocorrelation functions we can extract effective viscosities and determine the critical micellization temperature and concentration. Moreover,the high EO/PO (arm-to-core) ratio of F127 and its polydispersity play a critical role in the high-temperature re-entrant liquid phase, due to decreasing solubility of PEO along with the dehydration of the PPO core. The microscopic viscoelastic moduli $G'(\omega)$ and $G''(\omega)$ help to determine these phase transitions and provide mechanical properties in the solid phase that are not readily accessible with standard multi-particle tracking techniques due to limited Brownian motion.
\end{abstract}

\maketitle


\section{\label{sec:level1}INTRODUCTION}

Triblock copolymers made of polyethylene oxide (PEO) and polypropylene oxide (PPO) are polymeric, non-ionic surfactants, commonly denoted as PEO-PPO-PEO. These symmetric ABA-type copolymers were introduced by BASF as poloxamers and are commercially also known as Pluronic\textsuperscript{\tiny\textregistered}, Symperonic\textsuperscript{\tiny\textregistered} or Kolliphor\textsuperscript{\tiny\textregistered}\,\cite{pitto2014pluronic,mortensen1992,wanka1990}. Unlike in small surfactant solutions, self-assembly of polymeric surfactants into microphase-separated structures is strongly temperature dependent\,\cite{kjellander1981water,cook1992pressure} and can show several phase transitions for a given volume fraction\,\cite{mortensen1992,hecht1995l3,brown1992,wanka1990,alexandridis1995,mortensen2008effects,malmsten1993effects,nolan1997light}. The type of phase formed depends both on the AB-block-size ratio and the poloxamer concentration. Just above their critical micelle temperature (CMT), most poloxamers with a PEO/PPO ratio larger than unity will form spherical micelles with a hydrophobic PPO core, which is stabilized by a solvated corona of hydrophilic PEO chains. At sufficiently high polymer concentrations and temperatures, these polymeric micelles crystallize into soft solids with cubic symmetry\,\cite{mortensen1992,hamley1998,eiser2000PRE}. 

Amongst different Pluronics\textsuperscript{\tiny\textregistered}, F127 (EO$_{100}$PO$_{65}$EO$_{100}$) stands out for its larger molecular weight (12600\,Da) and PEO/PPO ratio \cite{mortensen1992,pitto2014pluronic}. Because of its thermally reversible gelation property at around body temperature for concentrations above $\sim$\,16 wt\%, it is particularly interesting for biomedical applications, such as drug delivery and administration\cite{dotivo2021immobilization,akash2015recent}, and artificial skin for treating burns \cite{schmolka1972artificial}, amongst others\cite{akash2015recent}. Pitto-Barry and Barry\,\cite{pitto2014pluronic} give an excellent overview of F127 in medical applications. Many of these applications require knowledge of the viscoelastic and flow properties of these aqueous F127 solutions. The micellization and gelation with corresponding temperatures and concentrations have been studied previously \cite{wanka1990, attwood1985micellar, zhou1988light}. In particular, Mortensen and co-workers have studied the solid phase of F127 solutions with Small-Angle Neutron Scattering (SANS)\cite{mortensen1992,mortensen2008effects,wanka1994phase} and cryo Transmission Electron Microscopy\,\cite{mortensen1995cryo}. However, less is known about the viscoelastic properties of their solutions, in particular, around the unimer to micellar liquid, and micellar liquid to  micellar solid phase transitions. 

In this paper, the aggregation behavior of F127 in aqueous solution is studied with Diffusing Wave Spectroscopy based microrheology (DWS-MR). This technique allows us to measure small changes in the viscoelastic properties of low-viscosity systems, such as aqueous Pluronic solutions at low temperatures, as well as their solid-to-liquid transition upon heating to $\sim 80\,^{\circ}$C, without any evaporation effects faced in classical rheology \cite{gadzekpo2025integrative}. Moreover, we can access the system's high-frequency response, thereby complementing low-frequency data from classical bulk rheology \cite{stoev2021bulk,can2025mechanically}. Following the experimental section, where we detail the operation principles of DWS-MR, we present our phase diagram and microrheology data. We show that our MR data provide an excellent extension to classical rheology measurements and deliver complementary insights into the internal dynamics and micellar re-arrangements. Further, we relate our DWS-MR data to the corresponding Small-Angle Neutron and X-ray Scattering (SANS\cite{mortensen1996structural} and SAXS\cite{Liu2022}) data reflecting the phase textures of the system. We conclude with an outlook for further studies.

\begin{figure}
\includegraphics[width=\columnwidth]{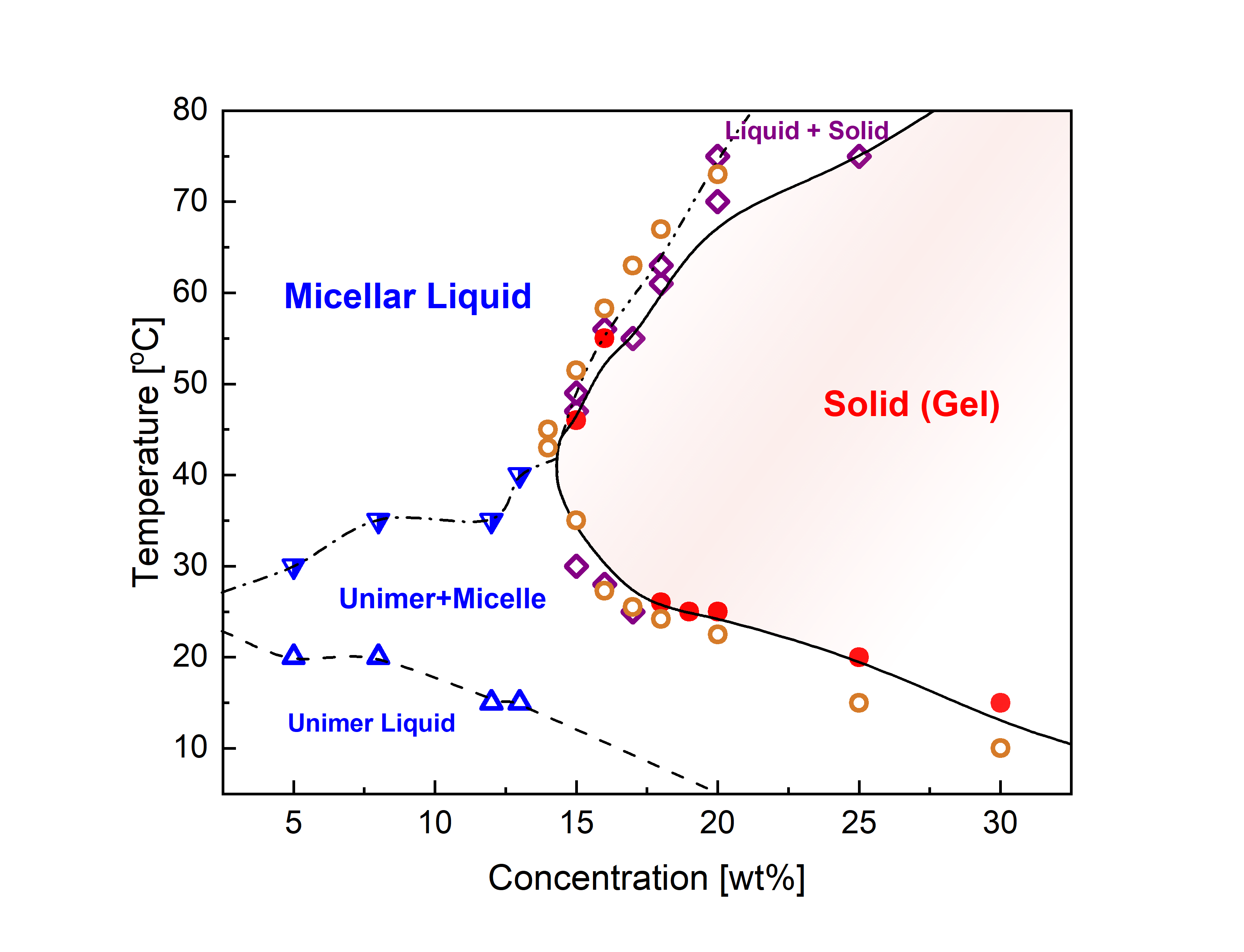}
    \caption{\label{fig:phase diagram}The phase diagram of F127 in deionized water. Orange open circles indicate the transition from a liquid to a gel and back to a liquid state, using visual observations upon heating the samples. All other symbols were obtained from DWS-MR measurements. The liquid-solid boundary was extracted from $g^{(2)}(\tau)$ curves (solid red circles and black line as guide to the eye; see also Fig.\,\ref{fig:ICF_MSD}), and $G'(\omega)$ data (violet diamonds). The blue upright triangles were obtained from viscosity measurements, which were derived from MSD curves, with the open triangles marking the CMT (dashed line), while the downwards pointing triangles along the dash-dotted line indicate the transition from a binary liquid of unimers and micelles to a purely micellar liquid. The dash-double-dotted line represents the transition from the solid gel and re-appearing liquid phase to a completely liquid micellar phase.}
\end{figure}

\section{\label{sec:level1}Experimental}

\subsection{Materials and Sample Preparation} 
Pluronic\textsuperscript{\tiny\textregistered}\, F127 was purchased from Sigma-Aldrich and used without further purification. The samples were prepared in 50\,mL centrifuge tubes filled with 30\,mL of Direct-Q\textsuperscript{\tiny\textregistered}-purified water. Appropriate amounts of dry F127 powder were gradually added to the tubes and dispersed on a magnetic stirrer set to around 1000\,rpm to facilitate dissolution. Samples with F127 concentrations above 15 wt\% were cooled in an ice bath to enable full dissolution. The F127 concentrations ranged from 5\,wt\% to 30\,wt\% and were confirmed by measuring the refractive index of the final concentrations with an Abbe refractometer. The fully dispersed samples were sealed with Teflon tape and Parafilm to prevent evaporation and stored at $4\,^{\circ}$C until further use.

\subsection{DWS-Microrheology}
All microrheology measurements were performed with a DWS RheoLab\texttrademark\, from LS Instruments AG (Fribourg, Switzerland). In DWS, we measure multiply scattered light generated by spherical polystyrene (PS) probe-particles. Due to the thermal diffusion of these probe particles, the multiply scattered laser light forms coherent light speckles that change in time. Here we use the instrument in forward-scattering mode \cite{xing2018microrheology}. The normalized intensity autocorrelation function $g^{(2)}(q,\tau)$ is related to the time dependent, scattered electric-field autocorrelation function $g^{(1)}(q,\tau)$ via the Siegert relation  $g^{(2)}(\tau)= 1+\beta |g^{(1)}(q,\tau)^2|$, where $\beta$ is a factor close to unity that reflects the point-like geometry of the avalanche detector, $\tau$ is the delay time, and $q$ is the scattering wavevector \cite{stoev2020role,mukhopadhyay2022amyloid}. For sufficiently high concentrations of non-interacting spherical particles, we can assume that the light itself behaves diffusive. In this case $g^{(1)}(q,\tau)$ can be derived, assuming that the scattered light performs a random walk \cite{pine1988DWS}. The expression of $g^{(1)}(q,\tau)$
directly contains the mean-squared displacements (MSDs) of the particles:
\begin{equation}\label{eqn:stokes-einstein}
    \langle\Delta r^2(\tau)\rangle = 6D\tau = \frac{k_B T}{\pi \eta a} \tau.
\end{equation}
Here, $D$ is the translational diffusion coefficient and $a$ is the radius of the diffusing tracer particles, while $\eta$ is the viscosity of the surrounding medium.
The thermal energy enters through the product of the Boltzmann constant $k_B$ and the temperature $T$ in Kelvin. 

We obtain the viscoelastic properties of the embedding environment by Laplace-Fourier transforming the MSDs of the probe-particles through the generalized Stokes-Einstein relation\cite{mason1995optical}:
\begin{equation}\label{eqn:GSER}
    G^*(\omega) = \frac{k_B T}{\pi a i \omega \langle \Delta \Tilde{r}^2(\omega)\rangle},
\end{equation}
where $G^*(\omega)$ is the complex shear modulus, $\langle \Delta \Tilde{r}^2(\omega)\rangle$ is the Laplace-Fourier transform of the MSD and $\omega$ is the angular frequency. The complex modulus $G^*(\omega)$ can be separated into the elastic $G'(\omega)$ and viscous $G''(\omega)$ moduli via $G^*(\omega) = G'(\omega)+iG''(\omega)$ using the Kramers-Kronig relations \cite{pine1990diffusing,scheffold2001diffusing,zakharov2006multispeckle}.

Maret and Wolf\cite{maret1987multiple} showed that the photon path-length distribution is related to the electric field auto-correlation function:
\begin{equation}\label{eqn:acf}
    g^{(1)}(q,\tau)=\int_{0}^{\infty} \mathrm{d}s P(s) \exp \left[ -\frac{1}{3}q^2 \langle\Delta r^2(\tau)\rangle \frac{s}{l^*}\right].
\end{equation}
Here, $P(s)$ is the probability distribution of the paths the light travels through the sample and $l^*$ is the mean-free path between successive scattering events. In the experiments, one needs to ensure that $l^*$ is appropriately adjusted, such that the path of the multiply scattered light is fully randomized as the photons pass through the cuvette. This requires a turbidity of minimum 70\%. Here we used 1$\times$1\,cm thick cuvettes and about 1\,wt\% of either 646\,nm or 720\,nm large PS particles were added to all samples measured with DWS. The two different probe sizes were used to verify that the results are not particle size dependent. Cuvettes and particles were provided by LS Instruments AG.

\section{RESULTS}

\textbf{Optical Observations.} The phase diagrams of various Pluronic\textsuperscript{\tiny\textregistered}\ (in the following referred to as Pluronic) solutions were studied previously \cite{wanka1994phase,zhou1994phase,malmsten1993effects,pitto2014pluronic}, demonstrating that their liquid-to-gel phase transition was caused by their self-assembly into spherical micelles. As Pluronics differ from batch to batch, we established the phase diagram of our system first. This was done by optical observation of F127 solutions with concentrations ranging from 5 to 30\,wt\%. For that, 10\,mL samples, sealed in transparent tubes, were placed in a temperature-controlled water bath. The samples were inspected by gently tumbling the tubes every $5\,^{\circ}$C step, heating the samples from ice-water to $80\,^{\circ}$C and subsequent cooling back to room temperature. In Fig.\,\ref{fig:phase diagram} we show the phase diagram obtained from our macroscopic observations of the liquid-to-solid transition measured for a heating ramp (orange open circles), as well as data extracted from DWS-MR data (triangles  and full circles). We note that the transition from a unimer to a binary liquid of unimers and micelles at the critical micellar temperature (dashed line in Fig.\,\ref{fig:phase diagram}) cannot be observed by visual inspection as the samples are completely transparent and fluid. The transition line indicating that all unimers have been converted into micelles, rendering the system a pure micellar liquid is indicted by the dash-dotted line. These transitions were extracted from DWS-MR measurements and will be discussed later.

The micellar solid region started to appear at F127 concentrations of $\sim$\,16\,wt\%. We observed that the transition from a liquid to a solid phase is relatively sharp and corresponds to a first-order phase transition, with a hardly measurable two-phase region, as reported for various other Pluronic systems \cite{wanka1990,malmsten1992self,mortensen1992,mortensen1996structural,eiser2000EPJE}. At high temperatures, the F127 samples showed a re-entrant liquefaction of the entire sample between 16 and $\sim$\,22\,wt\%, while above that concentration the entire sample remained predominantly solid with only a very small fraction becoming liquid at $80\,^{\circ}$C. For example, the 19\,wt\% sample showed a purely solid behavior between 25 and $65\,^{\circ}$C, while 30\,wt\% samples remained solid from around $15\,^{\circ}$C up to $80\,^{\circ}$C.
\begin{figure*}
    \centering \includegraphics[width=0.7\textwidth]{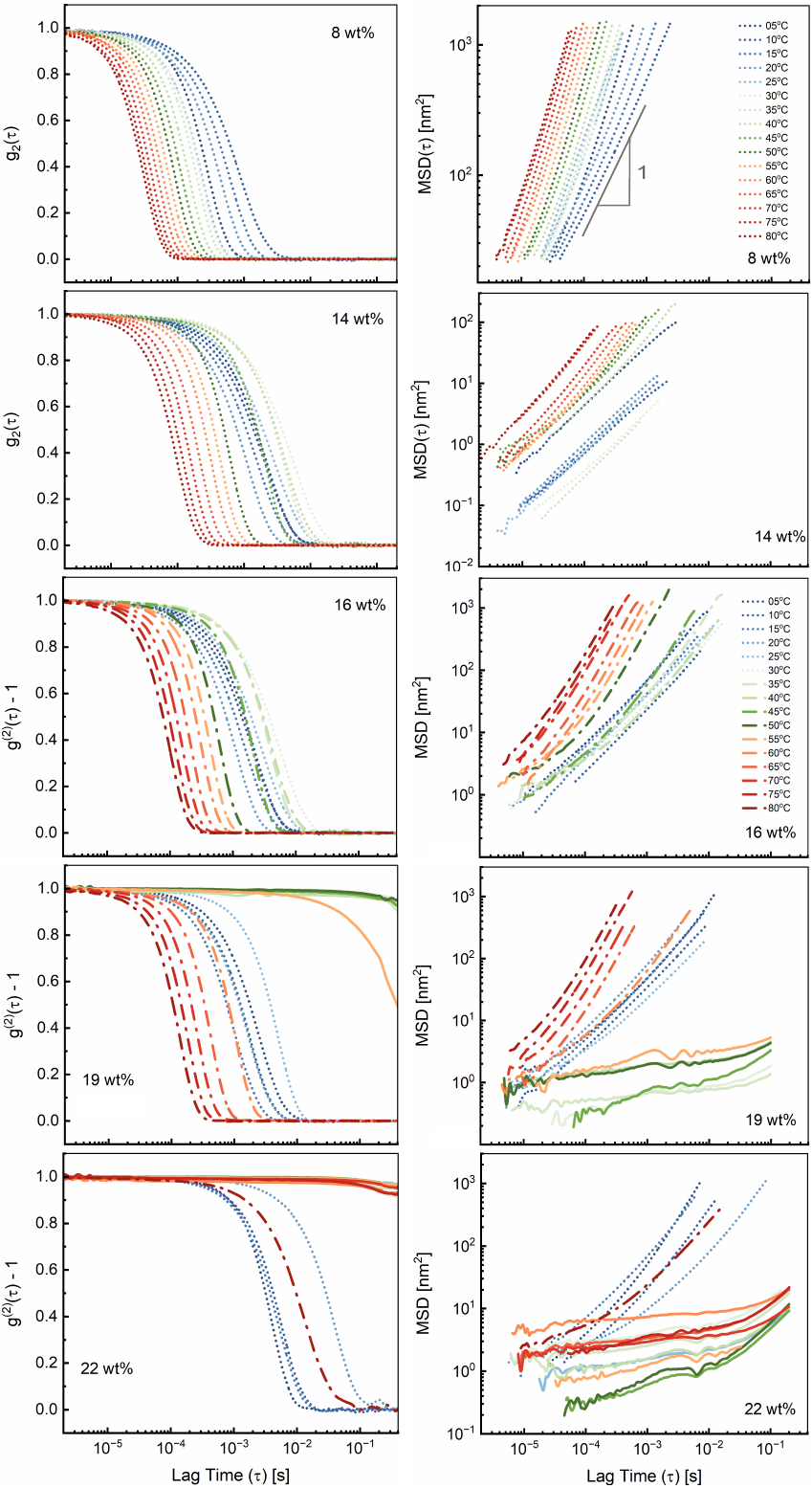}
    \caption{\label{fig:ICF_MSD}Intensity autocorrelation functions, $g^{(2)}(\tau)$, and the corresponding MSDs for different F127 concentrations that were measured for temperatures decreasing from 80$\,^{\circ}$C (red lines) to 5$\,^{\circ}$C (blue lines). The dash-doted lines represent the re-entrant liquid phase at high temperatures, the solid lines a solid gel phase and the dotted lines the liquid phases observed at low temperatures.}
\end{figure*}

\textbf{DWS Measurements.} We first identified the liquid and solid gel phases from $g^{(2)}(\tau)$ and the corresponding MSD curves for similar temperatures and concentrations we used in our optical studies. In Fig.\,\ref{fig:ICF_MSD} we show the typical changes of the intensity-autocorrelation functions of 8, 14, 16, 19 and 22\,wt\% samples measured in a cooling ramp starting at 80$\,^{\circ}$C. Almost identical results were obtained in a heating ramp. Similar to our optical experiments, the $g^{(2)}(\tau)$ curves of the 8 and 14\,wt\% samples decay completely for all temperatures (dotted lines) reflecting the fact that they remain fully liquid at all temperatures. For the 8\,wt\% sample, we observe that the times at which the curves decay to half ($\tau_{0.5}$) do not follow the monotonic temperature dependence of pure water. They are shortest at 80$\,^{\circ}$C followed by an increase in $\tau_{0.5}$ until 10$\,^{\circ}$C to then decrease again significantly at 5$\,^{\circ}$C. Moreover, the 25 and 30$\,^{\circ}$C curves are almost on top of each other and with slightly different slopes. Following previous results for solutions of Pluronic F68 obtained with classical rheology \cite{wu2006saltPluronics},
we interpret this non-monotonic behavior as the transition from the region in which unimers and micelles co-exist to a pure micellar dispersion. Conversely, the sudden increase in $\tau_{0.5}$ between the 5$\,^{\circ}$C and 10$\,^{\circ}$C curves marks the transition from the pure unimer to the coexistence region of unimers and micelles in solution. We also note that MSDs of the 8\,wt\% and 14\,wt\% samples do not display a purely diffusive behavior, as one would expect from a Newtonian liquid, but rather show a varying slope (see Fig.\,\ref{fig:ICF_MSD}). For comparison, the MSD of a suspension of monodisperse particles displays a linear time dependence with a slope of 1. SANS measurements by Mortenson and Talmon showed that the intensity  of the scattering function $S(q)$ of a 5\,wt\% F127 sample decreased considerably at small scattering wavevectors $q$, which reflects weak inter-micellar interactions due to hard-sphere repulsion even at lower temperatures and concentrations \cite{mortensen1995cryo}. This leads to an increased viscosity of the F127 suspension, which in turn slows down the diffusion of the tracer colloids. But to understand what causes the varying slopes in the extracted MSD curves, we note that Pluronics do not show a sharp unimer-to-micelle transition as observed in solutions of small-molecule surfactants. Already in the unimer region, at temperatures of a few degrees Celsius, the PPO blocks can form a collapsed core or micelles with varying number of Pluronic chains. These are highly short-lived and dynamic. Therefore, we expect to have a zoo of differently sized unimer-micelles and fully formed micelles even below the CMT or CMC, as observed in surface-tension measurements \cite{wanka1990,linse1993micellization,alexandridis1995} and simulations \cite{erika2004tripolymer}. We note that also polydispersity in the polymer chain lengths may add to the variation in the MSDs, as we will see later in the discussion. Size polydispersity of the tracer colloids remains negligible in our system.

The MSDs of the 14\,wt\% sample show a clear change from a more diffusive liquid state to a sub-diffusive one between 25$\,^{\circ}$C and 30$\,^{\circ}$C, which becomes more pronounced in the MSDs of the 16\,wt\% sample. Similarly to lower concentrations, the $g^{(2)}(\tau)$ curves shift to shorter decay times for increasing temperatures but then jump to larger decay times at 30$\,^{\circ}$C, followed by another jump at 40$\,^{\circ}$C. Within the narrow temperature range of 40$\,^{\circ}$C to 50$\,^{\circ}$C the sample shows solid-like behavior, as we will see from the corresponding $G'(\omega)$ and $G''(\omega)$ curves shown later. At even higher temperatures the samples fluidize again. 

The solid behavior is also reflected in the plateauing of the MSD curves. The temperature range, in which we observe the solid phase, widens with increasing F127 concentration, as seen form the 19\,wt\% and 22\,wt\% samples in Fig.\,\ref{fig:ICF_MSD}. This is reflected in the sudden jump of the half-times $\tau_{0.5}$ of the intensity-autocorrelation curves to values that are larger than 0.1\,s, which is close to the limit our instrument can resolve. The solid phase behavior is also expressed by the relative flatness of the corresponding MSDs in this temperature range. 

Above roughly 55$\,^{\circ}$C, the solid phases of the 19\,wt\% and 22\,wt\% samples melt, rendering them completely fluid again. We note that all samples display much shorter decay times in the $g^{(2)}(\tau)$ curves at the highest temperatures than those observed for the fluid phases at 5$\,^{\circ}$C, with the exception of the 22\,wt\% sample. Comparing the half-times at 80$\,^{\circ}$C of all concentrations shown in Fig.\,\ref{fig:ICF_MSD}, we see an interesting trend: while the 8\,wt\% sample shows the shortest
$\tau_{0.5}$, the 14\,wt\%, 16\,wt\% and 19\,wt\% display a slower but similar half time. Only the 22\,wt\% sample shows a significant increase by almost two orders of magnitude. It is possible that the additional increase in concentration causes stronger interactions between the hot micelles. From the long-time MSD-values we also extracted the approximate zero-shear viscosities for the liquid phases of the different F127 concentrations. These reflect nicely the onset of micellization (CMT) and the moment most chains are converted into micelles. These data also show similarities to the halftime plots.

The re-entrant liquefaction at high temperatures can be explained by the fact that PEO chains in water have a lower critical solution point at around the boiling temperature of water \cite{saeki1976upper,kjellander1981water,alexandridis1995,mortensen1995cryo}. Various experimental studies suggest that in the temperature range of 30$\,^{\circ}$C to 50$\,^{\circ}$C, the size of the F127 micelles does not change significantly \cite{kjellander1981water,wanka1990,mortensen1992,mortensen1995cryo}. However, at higher temperatures two effects occur. First, the solubility of the PEO chains starts to decrease leading to a weak shrinking of the water-soluble corona formed by the relatively long PEO chains in F127. Hence, even though enough micelles are formed to reach the volume fraction for a space-filling simple cubic crystal, their solid melts again upon further heating due to the effective shrinkage of their sizes. Second, several Pluronics undergo a transition from spherical micelles to ellipsoid or worm-like shape with an increased aggregation number of chains per micelle. This may be supported by the fact that as-received Pluronic contains a considerable amount of shorter diblock copolymers that may change the spontaneous curvature of the micelles. However, we do not think that the sphere to rod-like transition is sufficient to explain the strong liquefaction at high temperatures. We conclude that the hypothesis of an increasingly shrinking corona of the micelles is in line with our observation of a re-entrant melting region, which disappears with concentrations larger than 22\,wt\%.

\begin{figure*}
\includegraphics{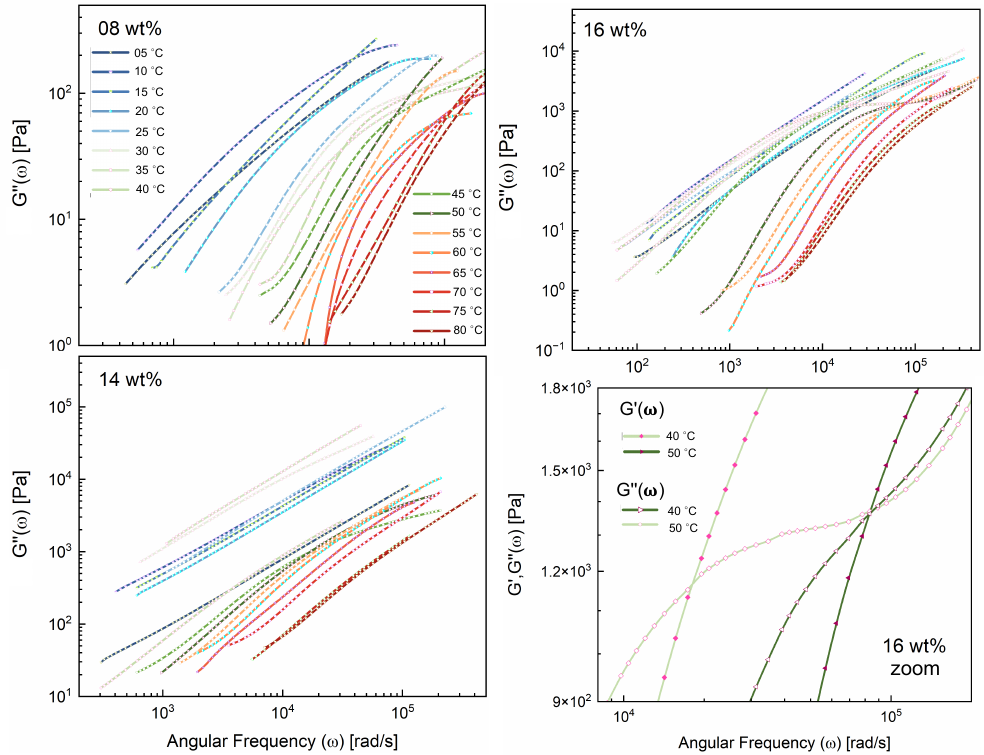}\caption{\label{fig:Gpp_comp1}Presentation of the loss moduli $G''(\omega)$ extracted from the MSDs of the 8\,wt\%, 14\,wt\% and 16\,wt\% samples presented in Fig.\ref{fig:ICF_MSD}. For all samples the elastic modulus $G'(\omega)$ was negligible, except for the 16\,wt\% sample, where we observe solid behavior only at 40$\,^{\circ}$C and 50$\,^{\circ}$C. At these temperatures $G'(\omega)$ exceeds the values of $G''(\omega)$. This is highlighted in the zoomed-in figure.}
\end{figure*}

\begin{figure*}
\includegraphics{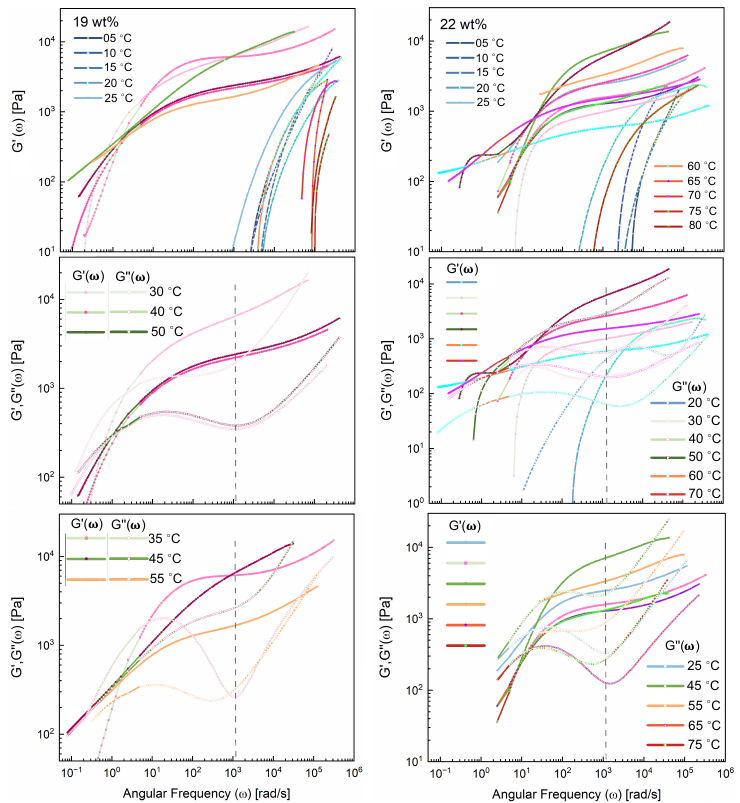}\caption{\label{fig:GpGpp_comp2}Presentation of $G'(\omega)$ and $G''(\omega)$ curves of the 19 and 22\,wt\% samples presented in Fig.\ref{fig:ICF_MSD}. In the top images the $G'(\omega)$ curves are plotted for all temperatures. For better visibility we plot both moduli in the lower two rows of images at even and odd temperatures in the solid phase.}
\end{figure*}

From our phase diagram and the large jump in the half-times of the autocorrelation functions, we know that the transition from a micellar liquid to a solid with cubic packing fraction of $\phi = 53\%$ occurs between 25$\,^{\circ}$C and 30$\,^{\circ}$C for the 19\,wt\% sample, and between 20$\,^{\circ}$C and 25$\,^{\circ}$C for the 22\,wt\% sample. This becomes also evident from the sudden rise of the elastic moduli to values above their corresponding loss moduli across many decades in frequencies. These liquid-to-solid transition temperatures and concentrations are in excellent agreement with oscillatory measurements by Wanka \textit{et al.} \cite{wanka1990}, who used classical bulk rheometers. However, the open structure of their sample geometry that allowed evaporation did not permit experiments above 40$\,^{\circ}$C and only in a limited frequency range. 

Further, the temperature range in which the 19\,wt\% and 22\,wt\% samples display a solid phase is in excellent agreement with the 2D-scattering ($S(q)$) images of a shear-aligned 20\,wt\% F127 sample taken with SANS \cite{mortensen1995cryo,mortensen1996structural}. The latter displayed a transition from a micellar liquid to a micellar solid with a predominantly powder-crystalline FCC structure at around 16$\,^{\circ}$C. Their shear-aligned sample showed the signature of a single crystal in the solid phase, which has been observed for a number of Pluronics \cite{eiser2000EPJE,eiser2000PRE,hamley1998}. And they also observe melting above around 65$\,^{\circ}$C, which again is similar to our observations. 

\begin{figure}[htbp]
    \centering
    \includegraphics[width=\columnwidth]{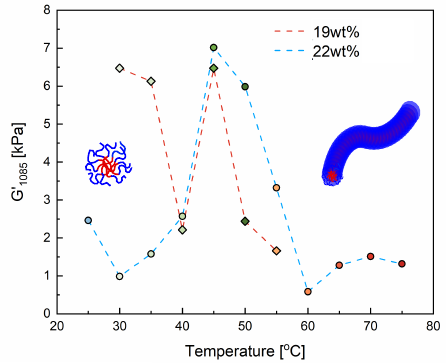}
    \caption{Plot of $G'(\omega)$ values for the 19 and 20\,wt\% F127 samples measured at 1085\,rad/s in the solid phase. The cartoons indicate that at lower temperatures the micelles are predominantly spherical while above 50$\,^{\circ}$C the micelles are believed to be more compact and worm-like \cite{mortensen1995cryo,Liu2022}.}
    \label{fig:Gp_values_solid}
\end{figure}

However, while most SANS and SAXS data show no significant change in the scattering functions at higher $q$ values when the concentration or temperature is increased, our $G'(\omega)$ data reveal an interesting change with temperature. Plotting the values of the elastic shear modulus measured for the 19\,wt\% at 1085\,rad/s, where $G''(\omega)$ is near its minimum in the solid phase, we observe an initial higher elasticity ($\sim$6500\,Pa) when we hit the liquid-to-solid transition (see Fig.\ref{fig:Gp_values_solid}). Upon further increase in temperature $G'(\omega)$ first drops to about $\sim$2200\,Pa at 40$\,^{\circ}$C and then increases back to $\sim$6500\,Pa at 45$\,^{\circ}$C, to then drop again to $\sim$2440\,Pa at 50$\,^{\circ}$C, before melting again at 60$\,^{\circ}$C. A similar behavior was observed for the 22\,wt\% sample, although with a lower elasticity value at the onset of solidification at low temperatures and the persistence of the solid behavior up to 75$\,^{\circ}$C.

\begin{table}[htbp]
\begin{ruledtabular}
\centering
\caption{\label{tab:table1}List of first crossover angular frequency $\omega_0$ between $G'(\omega)$ and $G''(\omega)$ and $G'$ values at 
$\omega_{b}=1,085$\,rad\,s$^{-1}$. }

\renewcommand{\arraystretch}{1.1}

\resizebox{\columnwidth}{!}{%
\begin{tabular}{lccccc}
\textbf{} & \textbf{T (°C)} & $\boldsymbol{\omega_0}$ (rad\,s$^{-1}$) 
& $\boldsymbol{G'(\omega_0)}$ (Pa) 
& $\boldsymbol{G'(\omega_b)}$ (Pa) 
& $\boldsymbol{\xi}$ (nm) \\ 
\midrule
\multicolumn{6}{l}{\textbf{19 wt\%}} \\ 
 & 25 & – & – & – & – \\
 & 30 & 2.5   & 988  & 6471 & 8.7 \\
 & 35 & 7.629 & 1791 & 6128 & 8.9 \\
 & 40 & 1667  & 106.4 & 2212 & 12.5 \\
 & 45 & 0.1   & – & 6471 & 8.8 \\
 & 50 & – & – & 2438 & 12.2 \\
 & 55 & – & – & 1661 & 14.0 \\
 & 60 & – & – & – & – \\ \hline
\addlinespace
\multicolumn{6}{l}{\textbf{22 wt\%}} \\
 & 15 & – & – & – & – \\
 & 20 & 3906 & 680  & 1924 & 4.3 \\
 & 25 & 13.9 & 689  & 2464 & 15.6 \\
 & 30 & 20   & 312  & 985  & 15.7 \\
 & 35 & 17   & 422  & 1579 & 15.8 \\
 & 40 & 14   & 632  & 2570 & 15.9 \\
 & 45 & 61   & 2228 & 7018 & 16.0 \\
 & 50 & 1.7  & 238  & 5985 & 16.1 \\
 & 55 & – & – & 3321 & 16.2 \\
 & 60 & – & – & 584  & 16.3 \\
 & 65 & 15.3 & 1282 & 1282 & 16.3 \\
 & 70 & – & – & 1509 & 16.4 \\
 & 75 & 15.3 & 369  & 1316 & – \\
 & 80 & 6510 & 509  & – & – \\
\end{tabular}}
\end{ruledtabular}
\end{table}

\section{\label{sec:level4}DISCUSSION}

DWS-microrheology data agree remarkably well with our simple, macroscopic phase-diagram measurements that rely on optical observations. Moreover, they provide us with additional information about the samples viscosities via the MSDs and the viscoelastic moduli. In particular, the latter show a remarkable temperature-dependent variations in the elastic modulus in the solid phase, while still reflecting characteristic length-scales that relate the elastic shear modulus measured near the minimum of $G''(\omega)$. The elastic bulk modulus of a thermally driven system, e.g. solutions of long, flexible polymers in the semi-dilute regime, can be described by the simple relation $G_{b} \sim k_BT/\xi^3$, where $\xi$ is the average entanglement length between the polymer chains. The measured characteristic interaction radius of the F127 micelles ranges around $10-12$\,nm using SANS \cite{mortensen1995cryo} and is around 10.1\,nm using SAXS \cite{Liu2022} for a 20\,wt\% F127 sample. The latter was obtained assuming a first scattering peak at $q_0= 0.38$\,nm for a FCC lattice. Using the simple expression above gives us characteristic elasticities of 2540 and 4260\,Pa at $T=45\,^{\circ}$C, respectively. These values are in the same range as those extracted from our measurements of $G'(\omega)$ at 1085\,rad/s, which are listed in Table\,\ref{tab:table1}. However, these simple arguments are insufficient to capture the strong variations in the $G'(\omega)$ values at different temperatures in the solid phase.

In a more recent publication, Mortensen \textit{et al.}\cite{mortensen2008effects} studied the effect of impurities in \textit{as-received} F127 samples on their crystalline structure in the solid gel phase. They report that 10\,wt\% to 25\,wt\% of the total mass of the samples are shorter polymers, such as PEO-PPO block copolymers with the same length as the PEO in F127 but half the length of the PPO block\cite{batsberg2004effects}. Moreover, they argue that the CMT of Pluronic solutions is strongly dependent on the PPO block length. This would contribute to the variation of the MSDs in our 8\,wt\% and 14\,wt\% samples at most temperatures and would result in a sharper CMT. Comparing the SANS data of solutions prepared with \textit{as-received} F127 with those of purified F127, they showed that both solutions exhibit similar scattering intensities in the colder liquid phase. At the micellar liquid-to-solid transition the purified samples went first from a disordered liquid to a solid with FCC symmetry, which switched to BCC structure at slightly higher temperatures and maintained that structure up to 80$\,^{\circ}$C. In contrast, the solid phase of the \textit{as-received} Pluronic displayed only FCC symmetry. These finding are in partial contrast with their earlier findings, which suggested that their samples melt above 65$\,^{\circ}$C.
\cite{mortensen1992,mortensen1995cryo} 

Our data presented here, as well as several other studies on F127 solutions find also melting at elevated temperatures \cite{Liu2022,jalaal2017rheology}. We hypothesize that the high amount of block copolymers is responsible for both the variations in the values of $G'(\omega)$ with temperature and partly responsible for the re-entrant melting between roughly 16\,wt\% and 25\,wt\%. Preliminary SAXS data \cite{Liu2022} performed with the same batch F127 presented here indeed show first Bragg-peaks of a few crystallites upon the transition to the solid phase, while a strong diffuse ring remains present in the background, which reflects the underlying disorder that may stem from the diblock copolymers. As Mortensen \textit{et al.} argued\cite{mortensen2008effects}, these shorter diblock copolymers most likely do not participate in the micelle formation of the triblocks but rather form smaller micelles at  higher temperatures, which may affect the elasticity of the system around 30-40$\,^{\circ}$C, while at elevated temperatures they may be incorporated in the F127 micelles as their corona starts to shrink and become elongated. We note that we observed a similar behavior in the viscoelastic properties in several concentrations and both in heating and cooling cycles. A possible explanation for the persistence of the crystalline phase up to 80$\,^{\circ}$C observed in some SANS and SAXS data may be due to the fact that the experiments were performed in 0.5\,mm capillaries that will have a larger surface-to-volume ratio and will thus represent a larger fraction of the crystalline phase forming at the flat container walls than the diffuse background of the bulk sample. In DWS, we use 1$\times$1\,cm thick cuvettes with a typical scattering length $l^*$ that is much larger than that of the probe particles, thus sampling the bulk behavior more efficiently. 

\section{\label{sec:level5}CONCLUSIONS}

We have demonstrated that DWS-based microrheology is an excellent tool to measure minute changes in the viscosity of aqueous suspensions of Pluronics, which allows us to determine the concentration-dependent critical micelle temperature of an otherwise completely transparent low-viscosity solution. At the same time, DWS-microrheology enables us to probe phase transitions marked by a steep change in the viscosity of a micellar liquid to a solid state as a function of temperature. Both observations are challenging or limited in parameter space when using classical macrorheology with cone-plate or Couette geometry. In addition, we can probe the equilibrium viscoelastic properties of many soft, complex systems over a wide frequency range, enabling us to also explore their high frequency or very local dynamics, which can be different from their bulk properties\cite{stoev2025microrheological}. From the measured relaxation curves, as well as the MSDs and viscoelastic moduli, we also obtain insight into the structure and dynamic re-arrangements of the samples, without having to do SANS, SAXS or other types of elaborate analysis, which can give us only structural information about the different phases of the system. To conclude, DWS-MR can be an excellent, fast tool for quickly testing a sample's viscosity, and thus concentration, as well as possible phase transitions in a relatively large parameter space. 

\begin{acknowledgments}
RT and EE are supported by the Research Council of Norway through its Centres of Excellence funding scheme (project no. 262644, PoreLab). The PhD-studentship of RT is funded by the RCN Project 'Sustainable Stable Ground' (grant no. 324486). IDS gratefully acknowledges funding from the Carl-Zeiss-Stiftung (Center SynGen) and the Karlsruhe Institute of Technology Excellence Strategy via the Young Investigator Group Preparation Program. RL received financial support from the Cambridge Trust and China Scholarship Council (CSC).
\end{acknowledgments}

\section{Author Contributions}
RT and XT are equally shared first authors of this article. RT, RL and IDS performed the experiments. XT, RT, IDS and EE analyzed the data. EE wrote the manuscript with inputs from all the authors.

\section {Data Availability Statement}
Data are available from the corresponding authors upon reasonable request.

\section{Conflicts of interest}
The authors have no conflicts of interest to disclose.

\nocite{*}
\bibliography{aipsamp}

@PREAMBLE{
 "\providecommand{\noopsort}[1]{}" 
 # "\providecommand{\singleletter}[1]{#1}%" 
}

@article{pitto2014pluronic,
  title={Pluronic{\textregistered} block-copolymers in medicine: from chemical and biological versatility to rationalisation and clinical advances},
  author={Pitto-Barry, Ana{\"\i}s and Barry, Nicolas PE},
  journal={Polym. Chem.},
  volume={5},
  number={10},
  pages={3291--3297},
  year={2014},
  publisher={Royal Society of Chemistry}
}

@article{mortensen1992,
  title={Phase Behaviour of Poly(ethylene oxide)-Poly(propylene oxide)-Poly(ethylene oxide) Triblock-Copolymer Dissolved in Water},
  author={Mortensen, Kell},
  journal={Europhys. Lett.},
  volume={19},
  number={7},
  pages={599--604},
  year={1992},
  publisher={Institute of Physics Publishing}
}

@article{wanka1990,
  title={The aggregation behavior of poly-(oxyethylene)-poly-(oxypropylene)-poly-(oxyethylene)-block-copolymers in aqueous solution},
  author={Wanka, George and Hoffman, H and Ulbricht, W},
  journal={Colloid Polym. Sci.},
  volume={268},
  pages={101–-117},
  year={1990},
  publisher={Springer}
}

@article{kjellander1981water,
  title={Water structure and changes in thermal stability of the system poly (ethylene oxide)--water},
  author={Kjellander, Roland and Florin, Ebba},
  journal={J. Chem. Soc., Faraday Trans. 1},
  volume={77},
  number={9},
  pages={2053--2077},
  year={1981},
  publisher={Royal Society of Chemistry}
}

@article{cook1992pressure,
  title={Pressure-induced crossover from good to poor solvent behavior for polyethylene oxide in water},
  author={Cook, Richard L and King Jr, HE and Peiffer, Dennis G},
  journal={Phys. Rev. Lett.},
  volume={69},
  number={21},
  pages={3072},
  year={1992},
  publisher={APS}
}

@article{hecht1995l3,
  title={L3 phase in a binary block copolymer/water system},
  author={Hecht, E and Mortensen, K and Hoffmann, H},
  journal={Macromolecules},
  volume={28},
  number={16},
  pages={5465--5476},
  year={1995},
  publisher={ACS Publications}
}

@article{brown1992,
  title={Triblock copolymers in aqueous solution studied by static and dynamic light scattering and oscillatory shear measurements: influence of relative block sizes},
  author={Brown, Wyn and Schillen, Karin and Hvidt, Soeren},
  journal={J. Phys. Chem.},
  volume={96},
  number={14},
  pages={6038–6044},
  year={1992},
  publisher={ACS Publications}
}

@article{alexandridis1995,
  title={Temperature Effects on Structural Properties of Pluronic P104 and F108 PEO-PPO-PEO Block Copolymer Solutions},
  author={Alexandridis, Paschalis and Nivaggioli, Thierry and Hatton, T. Alan},
  journal={Langmuir},
  volume={11},
  number={5},
  pages={1468--1476},
  year={1995},
  publisher={ACS Publications}
}

@article{mortensen2008effects,
  title={Effects of PEO- PPO diblock impurities on the cubic structure of aqueous PEO- PPO- PEO pluronics micelles: fcc and bcc ordered structures in F127},
  author={Mortensen, Kell and Batsberg, Walther and Hvidt, S{\o}ren},
  journal={Macromolecules},
  volume={41},
  number={5},
  pages={1720--1727},
  year={2008},
  publisher={ACS Publications}
}

@article{malmsten1993effects,
  title={Effects of homopolymers on the gel formation in aqueous block copolymer solutions},
  author={Malmsten, Martin and Lindman, Bjoern},
  journal={Macromolecules},
  volume={26},
  number={6},
  pages={1282--1286},
  year={1993},
  publisher={ACS Publications}
}

@article{nolan1997light,
  title={Light scattering study on the effect of polymer composition on the structural properties of PEO--PPO--PEO micelles},
  author={Nolan, Stuart L and Phillips, Ronald J and Cotts, Patricia M and Dungan, Stephanie R},
  journal={J. Colloid Interface Sci.},
  volume={191},
  number={2},
  pages={291--302},
  year={1997},
  publisher={Elsevier}
}

@article{hamley1998,
  title={Shear-Induced Orientational Transitions in the Body-Centered Cubic Phase of a Diblock Copolymer Gel},
  author={Hamley, Ian W. and Pople, J. A. and Fairclough, J. P. A. and Ryan, A. J. and Booth, C. and Yang, Y.-W.},
  journal={Macromolecules},
  volume={31},
  number={12},
  pages={3906–3911},
  year={1998},
  publisher={ACS Publications}
}

@article{eiser2000PRE,
  title={Nonhomogeneous textures and banded flow in a soft cubic phase under shear},
  author={Eiser, Erika and Molino, Francois and Porte, Gregoire and Diat, Olivier},
  journal={Phys. Rev. E},
  volume={61},
  number={6},
  pages={6759--6764},
  year={2000},
  publisher={American Physical Society}
}

@article{dotivo2021immobilization,
  title={Immobilization of PR4A3 enzyme in pluronic F127 polymeric micelles against colorectal adenocarcinoma cells and increase of in vitro bioavailability},
  author={Dotivo, Natielle Cachoeira and Rezende, Rachel Passos and Pessoa, Tharcilla Braz Alves and Salay, Luiz Carlos and Huachaca, N{\'e}lida Simona Mar{\'\i}n and Romano, Carla Cristina and Marques, Eric de Lima Silva and Costa, Moara Silva and de Moura, Suzana Rodrigues and Pirovani, Carlos Priminho and others},
  journal={Int. J. Biol. Macromol.},
  volume={166},
  pages={1238--1245},
  year={2021},
  publisher={Elsevier}
}

@article{akash2015recent,
  title={Recent progress in biomedical applications of Pluronic (PF127): Pharmaceutical perspectives},
  author={Akash, Muhammad Sajid Hamid and Rehman, Kanwal},
  journal={J. Control. Release.},
  volume={209},
  pages={120--138},
  year={2015},
  publisher={Elsevier}
}

@article{schmolka1972artificial,
  title={Artificial skin I. Preparation and properties of pluronic F-127 gels for treatment of burns},
  author={Schmolka, Irving R},
  journal={J. Biomed. Mater. Res.},
  volume={6},
  number={6},
  pages={571--582},
  year={1972},
  publisher={Wiley Online Library}
}

@article{attwood1985micellar,
  title={The micellar properties of the poly (oxyethylene)-poly (oxypropylene) copolymer Pluronic F127 in water and electrolyte solution},
  author={Attwood, D and Collett, JH and Tait, CJ},
  journal={Int. J. Pharm.},
  volume={26},
  number={1-2},
  pages={25--33},
  year={1985},
  publisher={Elsevier}
}

@article{zhou1988light,
  title={Light-scattering study on the association behavior of triblock polymers of ethylene oxide and propylene oxide in aqueous solution},
  author={Zhou, Zukang and Chu, Benjamin},
  journal={J. Colloid Interface Sci.},
  volume={126},
  number={1},
  pages={171--180},
  year={1988},
  publisher={Elsevier}
}

@article{wanka1994phase,
  title={Phase diagrams and aggregation behavior of poly (oxyethylene)-poly (oxypropylene)-poly (oxyethylene) triblock copolymers in aqueous solutions},
  author={Wanka, Gerd and Hoffmann, Heinz and Ulbricht, Werner},
  journal={Macromolecules},
  volume={27},
  number={15},
  pages={4145--4159},
  year={1994},
  publisher={ACS Publications}
}

@article{mortensen1995cryo,
  title={Cryo-TEM and SANS microstructural study of pluronic polymer solutions},
  author={Mortensen, Kell and Talmon, Yeshayahu},
  journal={Macromolecules},
  volume={28},
  number={26},
  pages={8829--8834},
  year={1995},
  publisher={ACS Publications}
}

@article{xing2018microrheology,
  title={Microrheology of DNA hydrogels},
  author={Xing, Zhongyang and Caciagli, Alessio and Cao, Tianyang and Stoev, Iliya and Zupkauskas, Mykolas and O’Neill, Thomas and Wenzel, Tobias and Lamboll, Robin and Liu, Dongsheng and Eiser, Erika},
  journal={Proc. Natl. Acad. Sci. U.S.A.},
  volume={115},
  number={32},
  pages={8137--8142},
  year={2018},
  publisher={National Academy of Sciences}
}

@article{pine1988DWS,
  title = {Diffusing wave spectroscopy},
  author = {Pine, D. J. and Weitz, D. A. and Chaikin, P. M. and Herbolzheimer, E.},
  journal = {Phys. Rev. Lett.},
  volume = {60},
  issue = {12},
  pages = {1134--1137},
  numpages = {0},
  year = {1988},
  month = {Mar},
  publisher = {American Physical Society},
  doi = {10.1103/PhysRevLett.60.1134},
  url = {https://link.aps.org/doi/10.1103/PhysRevLett.60.1134}
}

@article{mason1995optical,
  title={Optical measurements of frequency-dependent linear viscoelastic moduli of complex fluids},
  author={Mason, Thomas G and Weitz, David A},
  journal={Phys. Rev. Lett.},
  volume={74},
  number={7},
  pages={1250},
  year={1995},
  publisher={APS}
}

@article{pine1990diffusing,
  title={Diffusing-wave spectroscopy: dynamic light scattering in the multiple scattering limit},
  author={Pine, Dave J and Weitz, Dave A and Zhu, JX and Herbolzheimer, Eric},
  journal={J. Phys. France},
  volume={51},
  number={18},
  pages={2101--2127},
  year={1990},
  publisher={Soci{\'e}t{\'e} Fran{\c{c}}aise de Physique}
}

@article{scheffold2001diffusing,
  title={Diffusing-wave spectroscopy of nonergodic media},
  author={Scheffold, F and Skipetrov, SE and Romer, S and Schurtenberger, P},
  journal={Phys. Rev. E},
  volume={63},
  number={6},
  pages={061404},
  year={2001},
  publisher={APS}
}

@article{zakharov2006multispeckle,
  title={Multispeckle diffusing-wave spectroscopy with a single-mode detection scheme},
  author={Zakharov, Pavel and Cardinaux, Fr{\'e}d{\'e}ric and Scheffold, Frank},
  journal={Phys. Rev. E},
  volume={73},
  number={1},
  pages={011413},
  year={2006},
  publisher={APS}
}

@article{maret1987multiple,
  title={Multiple light scattering from disordered media. The effect of Brownian motion of scatterers},
  author={Maret, G and Wolf, PE},
  journal={Z. Phys. B: Condens. Matter},
  volume={65},
  number={4},
  pages={409--413},
  year={1987},
  publisher={Springer}
}

@article{zhou1994phase,
  title={Phase behavior and association properties of poly (oxypropylene)-poly (oxyethylene)-poly (oxypropylene) triblock copolymer in aqueous solution},
  author={Zhou, Zukang and Chu, Benjamin},
  journal={Macromolecules},
  volume={27},
  number={8},
  pages={2025--2033},
  year={1994},
  publisher={ACS Publications}
}

@article{malmsten1992self,
  title={Self-assembly in aqueous block copolymer solutions},
  author={Malmsten, Martin and Lindman, Bjoern},
  journal={Macromolecules},
  volume={25},
  number={20},
  pages={5440--5445},
  year={1992},
  publisher={ACS Publications}
}

@article{mortensen1996structural,
  title={Structural studies of aqueous solutions of PEO-PPO-PEO triblock copolymers, their micellar aggregates and mesophases; a small-angle neutron scattering study},
  author={Mortensen, Kell},
  journal={J. Phys. Condens. Matter},
  volume={8},
  number={25A},
  pages={A103},
  year={1996},
  publisher={IOP Publishing}
}

@article{eiser2000EPJE,
  title={Correlation between the viscoelastic properties of a soft crystal and its microstructure},
  author={Eiser, Erika and Molino, Francois and Porte, Gregoire},
  journal={Eur. Phys. J. E},
  volume={2},
  pages={39--46},
  year={2000},
  publisher={Springer-Verlag 2000}
}

@article{wu2006saltPluronics,
  title={Effect of salt on the phase behaviour of F68 triblock PEO/PPO/PEO copolymer},
  author={Wu, Yu Ling and Sprik, Rudolf and Poon, Wilson C. K. and Eiser, Erika},
  journal={J. Phys.: Condens. Matter},
  volume={18},
  number={19},
  pages={4461--4470},
  year={2006},
  publisher={Institute of Physics Publishing}
}

@article{linse1993micellization,
  title={Micellization of poly (ethylene oxide)-poly (propylene oxide) block copolymers in aqueous solution},
  author={Linse, Per},
  journal={Macromolecules},
  volume={26},
  number={17},
  pages={4437--4449},
  year={1993},
  publisher={ACS Publications}
}

@article{erika2004tripolymer,
  title={Simulation study of intra- and intermicellar ordering in triblock-copolymer systems},
  author={Wijmans, Chris M. and  Eiser, Erika  and Frenkel, Daan},
  journal={J. Chem. Phys.},
  volume={120},
  number={12},
  pages={5839–5848},
  year={2004},
  publisher={American Institute of Physics}
}

@article{saeki1976upper,
  title={Upper and lower critical solution temperatures in poly (ethylene glycol) solutions},
  author={Saeki, Susumu and Kuwahara, Nobuhiro and Nakata, Mitsuo and Kaneko, Motozo},
  journal={Polymer},
  volume={17},
  number={8},
  pages={685--689},
  year={1976},
  publisher={Elsevier}
}

@PHDTHESIS{Liu2022,
   author       = "R. Liu", 
   title        = "Designing Biocompatible Functional Hydrogels: Experiments and Simulations", 
   school       = "University of Cambridge", 
   year         = "2022", 
   type         = "{Ph.D.} thesis", 
   address      = "Cavendish Laboratory, J. J. Thomson Avenue, Cambridge CB3 0HE, U.K.", 
   month        = "November", 
   note         = "", 
}

@article{batsberg2004effects,
  title={Effects of poloxamer inhomogeneities on micellization in water},
  author={Batsberg, Walther and Ndoni, Sokol and Trandum, Christa and Hvidt, S{\o}ren},
  journal={Macromolecules},
  volume={37},
  number={8},
  pages={2965--2971},
  year={2004},
  publisher={ACS Publications}
}

@article{jalaal2017rheology,
  title={On the rheology of Pluronic F127 aqueous solutions},
  author={Jalaal, Maziyar and Cottrell, Graeme and Balmforth, Neil and Stoeber, Boris},
  journal={J. Rheol},
  volume={61},
  number={1},
  pages={139--146},
  year={2017},
  publisher={AIP Publishing}
}

@article{stoev2025microrheological,
  title={Microrheological characterisation of clay-PEO nanocomposites},
  author={Stoev, Iliya D and Mukhopadhyay, Anasua and Tammen, Rene and Eiser, Erika},
  journal={Rheol. Acta},
  volume={64},
  number={4},
  pages={241--253},
  year={2025},
  publisher={Springer}
}

@article{stoev2020role,
  title={On the role of flexibility in linker-mediated DNA hydrogels},
  author={Stoev, Iliya D and Cao, Tianyang and Caciagli, Alessio and Yu, Jiaming and Ness, Christopher and Liu, Ren and Ghosh, Rini and O’Neill, Thomas and Liu, Dongsheng and Eiser, Erika},
  journal={Soft Matter},
  volume={16},
  number={4},
  pages={990--1001},
  year={2020},
  publisher={Royal Society of Chemistry}
}

@article{stoev2021bulk,
  title={Bulk rheology of sticky DNA-functionalized emulsions},
  author={Stoev, Iliya D and Caciagli, Alessio and Mukhopadhyay, Anasua and Ness, Christopher and Eiser, Erika},
  journal={Phys. Rev. E},
  volume={104},
  number={5},
  pages={054602},
  year={2021},
  publisher={APS}
}

@article{can2025mechanically,
  title={Mechanically tunable DNA hydrogels as prospective biosensing modules},
  author={Can, Asya E and Ali, Abdul WU and Oelschlaeger, Claude and Willenbacher, Norbert and Stoev, Iliya D},
  journal={Macromol. Rapid Commun.},
  volume={46},
  number={23},
  pages={2500149},
  year={2025},
  publisher={Wiley Online Library}
}

@article{mukhopadhyay2022amyloid,
  title={Amyloid-like aggregation in native protein and its suppression in the bio-conjugated counterpart},
  author={Mukhopadhyay, Anasua and Stoev, Iliya D and King, David A and Sharma, Kamendra P and Eiser, Erika},
  journal={Front. Phys.},
  volume={10},
  pages={924864},
  year={2022},
  publisher={Frontiers Media SA}
}

@article{gadzekpo2025integrative,
  title={Integrative Approaches for DNA Sequence-Controlled Functional Materials},
  author={Gadzekpo, Aaron and Oprzeska-Zingrebe, Ewa Anna and Kozlowska, Mariana and Hilbert, Lennart and Stoev, Iliya D},
  journal={Adv. Funct. Mater.},
  pages={e19573},
  year={2025},
  publisher={Wiley Online Library}
}

\end{document}